\documentstyle [12pt,amssymb]{amsart}
\newtheorem{thm}{Theorem}
\newtheorem{cor}{Corollary}

\newtheorem{lem}{Lemma}

\begin{document}
\title{Modular Invariance and Characteristic Numbers}
\author{ Kefeng Liu}
\date {}
\maketitle

\begin{abstract}
We prove that a general miraculous cancellation formula, the divisibility of
certain characteristic numbers and some other topological results are
consequences of the modular invariance of elliptic operators on loop space.
\end{abstract}
\section{Motivations}
In [AW], a gravitational anomaly cancellation formula was derived from direct
computations. See also [GS] and [GSW]. This is essentially a formula relating
the $L$-class to the $\hat{A}$-class and a twisted $\hat{ A}$-class of a
$12$-dimensional manifold. More precisely, let $M$ be a smooth manifold of
dimension $12$, then the miraculous cancellation formula is
$$L(M)=8\hat{A}(M,T)-32\hat{A}(M)$$ where $T=TM$ denotes the tangent bundle of
$M$ and the equality holds at the top degree of each differential form. Here
recall that, if we use $\{ \pm x_j\}$ to denote the formal Chern roots of
$TM\otimes C$, then
$$L(M)=\prod_j\frac{x_j}{\mbox{tanh}x_j/2},\
\hat{A}(M)=\prod_j\frac{x_j/2}{\mbox{sinh}x_j/2},\ \mbox{and}$$
$$\hat{A}(M, T)=\hat{A}(M)\mbox{ch}T\ \mbox{with} \ \ \mbox{ch}T=\sum_j
e^{x_j}+e^{-x_j}.$$

By using this formula, Zhang [Z] derived an analytical version of Ochanine's
Rochlin congruence formula for $12$-dimensional manifolds. He achieved this by
considering the adiabatic limit of the $\eta$-invariant of a circle bundle over
a characteristic submanifold of $M$. In [LW], it was shown that a general
formula of such type implies that the generalized Rochlin invariant is a
spectral invariant.

On the other hand, Hirzebruch [H] and Landweber [L] used elliptic genus to
prove Ochanine's result: the signature of an $8k+4$-dimensional compact smooth
spin manifold is divisible by $16$. Actually Hirzebruch derived a general
formula in [H] relating the signature to the indices of certain twisted Dirac
operators of a compact smooth spin manifold of any dimension.

In this paper, using the modular invariance of certain elliptic operators on
loop space or the general elliptic genera, we derive a more general miraculous
cancellation formula about the characteristic classes of a smooth manifold and
a real vector bundle on it. This formula includes all of the above results and
are much more explicit. It is actually a formula of differential forms.
Combining with the index formula of [APS] for manifold with boundary and the
holonomy formula of determinant line bundle, we are able to show that several
seemingly unrelated topological results (in Corollaries 1-6) are connected
together by this general cancellation formula.

In the following we will state the results in Section 2 and prove them in
Section 3. In Section 4 we discuss some generalizations of the cancellation
formula. Also given in this section is a formula relating the generalized
Rochlin invariant to the holonomies of certain determinant line bundles.
Section 5 contains general discussions about other applications of modular
invariance in topology and some comments on the relationship between loop
space, double loop space and cohomology theory.

Our main results in this paper grew out of discussions with W. Zhang. The main
idea is essentially due to Hirzebruch [H] and Landweber [L]. My sole
contribution is to emphsize the role of modular invariance which reflects many
beautiful properties of the topology of manifold and its loop space. Note that
we have shown in [Liu1] that modular invariance also implies the rigidity of
many naturally derived elliptic operators from loop space. In a joint paper
with W. Zhang [LZ], combining the results of [Z] and the cancellation formula
in this paper, we will derive the  analytical expressions of the Finashin's
invariant [F] and some other topological invariants.
\section{Results}
Let $M$ be a dimension $8k+4$ smooth manifold and $V$ be a rank $2l$ real
vector bundle on $M$. We introduce two elements in $K(M)[[q^{\frac{1}{2}}]]$
which consists of formal power series in $q$ with coefficients in the $K$-group
of $M$.
\begin{eqnarray*}
\Theta_1(M,V)&=& \otimes_{n=1}^\infty
S_{q^n}(TM-\mbox{dim}M)\otimes_{m=1}^\infty\Lambda_{q^m}(V-\mbox{dim}V),\\
\Theta_2(M,V)&=& \otimes_{n=1}^\infty
S_{q^n}(TM-\mbox{dim}M)\otimes_{m=1}^\infty\Lambda_{-q^{m-\frac{1}{2}}}(V-\mbox{dim}V)
\end{eqnarray*}
where $q=e^{2\pi i\tau}$ with $\tau\in H$, the upper half plane, is a
parameter. Recall that for an indeterminant $t$,
$$\Lambda_t(V)=1+tV+t^2\Lambda^2V+\cdots,\ S_t(V)=1+tV+t^2S^2V+\cdots$$ are two
operations in $K(M)[[t]]$. They have relations
$$S_t(V)=\frac{1}{\Lambda_{-t}(V)},\ \Lambda_t(V-W)=\Lambda_t(V)S_{-t}(W).$$
We can formally expand $\Theta_1(M,V)$ and $\Theta_2(M,V)$ into Fourier series
in $q$
\begin{eqnarray*}
\Theta_1(M,V)&=&A_0+A_1q+\cdots,\\
\Theta_2(M,V)&=&B_0+B_1q^{\frac{1}{2}}+\cdots
\end{eqnarray*}
where the $A_j$'s and $ B_j$'s are elements in $K(M)$. Let $\{ \pm 2\pi iy_j\}$
be the formal Chern roots of $V\otimes C$. If $V$ is spin and $\triangle(V)$ is
the spinor bundle of $V$, one knows that the Chern character of $\triangle(V)$
is given by
$$\mbox{ch}\triangle(V)=\prod^l_{j=1}(e^{\pi iy_j}+e^{-\pi iy_j}).$$
In the following, we do not assume that $V$ is spin, but still formally use
$\mbox{ch}\triangle(V)$ for the short hand notation of $\prod^l_{j=1}(e^{\pi
iy_j}+e^{-\pi iy_j})$ which is a well-defined cohomology class on $M$. Let
$p_1$ denote the first Pontryagin class. Our main results include the following
theorem:
\begin{thm} If $p_1(V)=p_1(M)$, then
$$\hat{A}(M)\mbox{ch}\triangle(V)=2^{l+2k+1}\sum^k_{j=0}2^{-6j}b_j$$ where the
$b_j$'s are integral linear combinations of the $\hat{A}(M)chB_j$'s.
\end{thm}
Here still we only take the top degree terms of both sides. More generally we
have that for any $A_j$ as in the Fourier expansion of $\Theta_1(M,V)$, the top
degree term of $$\hat{A}(M)\mbox{ch}\triangle(V)\mbox{ch} A_j$$ is an integral
linear combination of the top degree terms of the $\hat{A}(M)\mbox{ch}B_j$'s.
All of the $b_j$'s can be computed explicitly, for example
$$b_0=-\hat{A}(M),\ b_1=\hat{A}(M)\mbox{ch}V+(24(2k+1)-2l)\hat{A}(M).$$

Take $V=TM$, then $\hat{A}(M)\mbox{ch}\triangle(V)=L(M)$. We get, at the top
degree, $$L(M)=2^3\sum_{j=0}^k 2^{6k-6j}b_j.$$ When dim$M=12$, one recovers the
miraculous cancellation formula of [AW].

For a dimension $8k$ manifold $M$ we have similar cancellation formula
$$L(M)=\sum_{j=0}^k 2^{6k-6j}b_j.$$One can also express the top degree of
$\hat{A}(M)$ in terms of those of the twisted $L$-classes.

Now assume $M$ is comact. As an easy corollary of Theorem 1, we have the
following
\begin{cor} If $M$ and $V$ are spin and $p_1(V)=p_1(M)$, then
$$\mbox{Ind}\ D\otimes\triangle(V)=2^{l+2k+1}\cdot\sum^k_{j=0}2^{-6j}b_j$$where
$D$ is the Dirac operator on $M$.
\end{cor}
The $b_j$'s are integral linear combinations of the $\mbox{Ind}D\otimes B_j$'s.
This gives
\begin{cor} If $M$ and $V$ are spin with $\mbox{dim}V\geq\mbox{dim}M$ and
$p_1(V)=p_1(M)$, then $\mbox{Ind}\ D\otimes\triangle(V)\equiv0\ (\mbox{mod}\
16),$ especially $\mbox{sign}(M)\equiv0\ (\mbox{mod}\ 16).$
\end{cor}

When $V=TM$ Corollaries 1 and 2 were derived in [H]. See also [L].

Recall the definition of the Rochlin invariant. Given a compact smooth
dimension $8k+3$ manifold $N$ with spin structure $w$, which is the spin
boundary of a spin manifold $M$ with spin structure $W$. The Rochlin invariant
$R(N, w)$ is defined to be $$R(N,w)=\mbox{sign}(M)(\mbox{mod}16).$$
Ochanine's theorem implies that $R(N,w)$ is well-defined. Let us formally write
$$b_j=\hat{A}(M)\mbox{ch}\beta_j$$ with $\beta_j$ the integral linear
combinations of the $B_j$'s. Another corollary of Theorem 1 is
\begin{cor}
The Rochlin invariant of $(N,w)$ is a spectral invariant of $N$ and is given by
the explicit formula
$$R(N,w)\equiv -\eta(\triangle)+\sum_{j=0}^k
2^{6k-6j+2}(\eta(\beta_j)+h_{\beta_j})(\mbox{mod}\ 16).$$
\end{cor}
Here $\eta(\triangle)$ and $\eta(\beta_j)$ are the $\eta$-invariant associated
to the signature operator $d_s=D\otimes \triangle(M)$ and $D\otimes \beta_j$
respectively. $h_{\beta_j}$ is the complex dimension of the kernal of the Dirac
operator on $N$ twisted by the restriction of $\beta_j$ to $N$. They depends on
the geometry of $(N,w)$. See Section 2 for their precise definitions.

For an oriented smooth compact manifold $M$ of dimension $8k+4$, let $F$ be a
submanifold which is the Poincare dual of the second Stiefel-Whitney class
$w_2(M)$. Denote by $F\cdot F$ the self-intersection of $F$. Let $D_F$ denote
the Dirac operator on $F$, $\beta_J^F$ be the restriction of $\beta_j$ to $F$.
In [LZ], combining the cancellation formula in Theorem 1 and the results in
[Z], we have proved the following analytic version of the generalized Rochlin
congruence formula of Finashin [F]:
\begin{cor}
 $$\frac{\mbox{sign}(M)+\mbox{sign}(F\cdot F)}{8}\equiv \sum_{j=0}^k
2^{6k-6j-1}(\eta(\beta_j)+h_{\beta_j})\ (\mbox{mod}\ 2).$$
\end{cor}
Here $\eta(\beta_j)$ and $h_{\beta_j}$ are the corresponding $\eta$-invariant
and the dimension of the kernal of certain twisted Dirac operator on $F$ by the
restriction of $\beta_j$ to $F$. The proof of Corollary 4 and some other more
general results in this direction can be found in [LZ].

For the divisibility of the twisted signature and the relation between the
generalized Rochlin invariant and the holonomies of determinant line bundles,
see Section 4.
\section{Proofs}

The proof of Theorem 1 needs the Jacobi theta-functions [Ch] which are
\begin{eqnarray*}
\theta(v,\tau)&=&2q^{\frac{1}{8}}\mbox{sin}\pi
v\prod_{j=1}^\infty(1-q^j)(1-e^{2\pi iv}q^j)(1-e^{-2\pi iv}q^j),\\
\theta_1(v,\tau)&=&2q^{\frac{1}{8}}\mbox{cos}\pi
v\prod_{j=1}^\infty(1-q^j)(1+e^{2\pi iv}q^j)(1+e^{-2\pi iv}q^j),\\
\theta_2(v,\tau)&=&\prod_{j=1}^\infty(1-q^j)(1-e^{2\pi
iv}q^{j-\frac{1}{2}})(1-e^{-2\pi iv}q^{j-\frac{1}
{2}}), \  \mbox{and}\\
\theta_3(v,\tau)&=&\prod_{j=1}^\infty(1-q^j)(1+e^{2\pi
iv}q^{j-\frac{1}{2}})(1+e^{-2\pi iv}q^{j-\frac{1}{2}}).
\end{eqnarray*}
These are holomorphic functions for $(v,\tau)\in C\times H$ where $C$ is the
complex plane and $H$ is the upper half plane. Up to some complex constants
they have the following remarkable transformation formulas
\begin{eqnarray*}
\theta(\frac{v}{\tau},-\frac{1}{\tau})=i\tau^{\frac{1}{2}}e^{-\frac{v^2}{\tau}}\theta(v,\tau)&,&
\theta(v,\tau+1)=\theta(v,\tau);\\
\theta_1(\frac{v}{\tau},-\frac{1}{\tau})=\tau^{\frac{1}{2}}e^{-\frac{v^2}{\tau}}\theta_2(v,\tau)&,&\theta_1(v,\tau+1)=\theta_1(v,\tau);\\
\theta_2(\frac{v}{\tau},-\frac{1}{\tau})=\tau^{\frac{1}{2}}e^{-\frac{v^2}{\tau}}\theta_1(v,\tau)&,&\theta_2(v,\tau+1)=\theta_3(v,\tau);\\
\theta_3(\frac{v}{\tau},-\frac{1}{\tau})=\tau^{\frac{1}{2}}e^{-\frac{v^2}{\tau}}\theta_3(v,\tau)&,&\theta_3(v,\tau+1)=\theta_2(v,\tau).\\
\end{eqnarray*}
These transformation formulas which are simple consequences of the Poisson
summation formula will be the key of our argument. Let
\begin{eqnarray*}
\Gamma_0(2)&=&\{ \left(\begin{array}{cc}a&b\\c&d\end{array}\right)\in SL_2(Z)|
c\equiv0\ (\mbox{mod}\ 2)\},\\
\Gamma^0(2)&=&\{ \left(\begin{array}{cc}a&b\\c&d\end{array}\right)\in
SL_2(Z)|b\equiv0\ (\mbox{mod}\ 2)\}
\end{eqnarray*}
 be two modular subgroups of $SL_2(Z)$. Let
$S=\left(\begin{array}{cc}0&-1\\1&0\end{array}\right),\
T=\left(\begin{array}{cc}1&1\\0&1\end{array}\right)$ be the two generators of
$SL_2(Z)$. Their actions on $H$ are given by
$$S:\ \tau\rightarrow -\frac{1}{\tau},\ \ T:\ \tau \rightarrow \tau+1.$$
Recall that a modular form over a modular subgroup $\Gamma$ is a holomorphic
function $f(\tau)$ on $H$ which, for any
$g=\left(\begin{array}{cc}a&b\\c&d\end{array}\right)\in \Gamma$, satisfies the
transformation formula
$$f(\frac{a\tau+b}{c\tau+d})=\chi(g)(c\tau+d)^kf(\tau)$$where $\chi:\
\Gamma\rightarrow C^*$ is a character of $\Gamma$ and $k$ is called the weight
of $f$. We also assume $f$ is holomorphic at $\tau=i\infty$.

Obviously, at $v=0$, $\theta_j(0,\tau)$ are modular forms of weight
$\frac{1}{2}$ and $$\theta'(0,\tau)=\frac{\partial}{\partial
v}\theta(v,\tau)|_{v=0}$$ is a modular form of weight $\frac{3}{2}$.

 With the notations as above we then have the following
\begin{lem}Assume $p_1(V)=p_1(M)$, then
$P_1(\tau)=\hat{A}(M)\mbox{ch}\triangle(V)\mbox{ch}\Theta_1(M,V)$ is a modular
form of weight $4k+2$ over $\Gamma_0(2)$;
$P_2(\tau)=\hat{A}(M)\mbox{ch}\Theta_2(M,V)$ is a modular form of weight $4k+2$
over $\Gamma^0(2)$.
\end{lem}
{\bf  Proof:} Let $\{ \pm 2\pi iy_\nu\}$ and $\{ \pm 2\pi ix_j\}$ be the
corresponding formal Chern roots of $V\otimes C$ and $TM\otimes C$. In terms of
the theta-functions, we have
\begin{eqnarray*}
P_1(\tau)&=&2^l(\prod_{j=1}^{4k+2}x_j\frac{\theta'(0,\tau)}{\theta(x_j,\tau)})\prod_{\nu=1}^l\frac{\theta_1(y_\nu,\tau)}{\theta_1(0,\tau)},\\
P_2(\tau)&=&(\prod_{j=1}^{4k+2}x_j\frac{\theta'(0,\tau)}{\theta(x_j,\tau)})\prod_{\nu=1}^l\frac{\theta_2(y_\nu,\tau)}{\theta_2(0,\tau)}.
\end{eqnarray*}
Apply the transformation formulas of the theta-functions, we have
$$P_1(-\frac{1}{\tau})=2^l\tau^{4k+2}P_2(\tau), \ P_1(\tau+1)=P_1(\tau)$$where
for the first equality, we need the condition $p_1(V)=p_1(M)$.

It is known that the generators of $\Gamma_0(2)$ are $T,\ ST^2ST$, while the
generators of $\Gamma^0(2)$ are $STS, \ T^2STS$ from which the lemma easily
follows.  $\spadesuit$

Write $\theta_j=\theta_j(0,\tau)$. We introduce some explicit modular forms
\begin{eqnarray*}
\delta_1(\tau)=\frac{1}{8}(\theta_2^4+\theta_3^4)&,& \
\varepsilon_1(\tau)=\frac{1}{16}\theta_2^4\theta_3^4,\\
\delta_2(\tau)=-\frac{1}{8}(\theta_1^4+\theta_3^4)&,& \
\varepsilon_2(\tau)=\frac{1}{16}\theta_1^4\theta_3^4.\\
\end{eqnarray*}
They have the following Fourier expansions in $q$
\begin{eqnarray*}
\delta_1(\tau)=\frac{1}{4}+6q+\cdots &,&
\varepsilon_1(\tau)=\frac{1}{16}-q+\cdots,\\
\delta_2(\tau)=-\frac{1}{8}-3q^{\frac{1}{2}}+\cdots &,&
\varepsilon_2(\tau)=q^{\frac{1}{2}}+\cdots,
\end{eqnarray*}
where $\cdots$ are the higher degree terms all of which are of integral
coefficients. Let $M(\Gamma)$ denote the rings of modular forms over $\Gamma$
with integral Fourier coefficients. We have
\begin{lem} $\delta_1, \delta_2$ are modular forms of weight $2$ and
$\varepsilon_1, \varepsilon_2$ are modular forms of weight $4$, and furthermore
$M(\Gamma^0(2))=Z[8\delta_2(\tau),\ \varepsilon_2(\tau)]$.
\end{lem}

The proof is quite easy. In fact the $\delta_2$ and $\varepsilon_2$ generate a
graded polynomial ring which has dimension $1+[\frac{k}{2}]$ in degree $2k$. On
the other hand one has a well-known upper bound for this dimension:
$1+\frac{k}{6}[SL_2(Z):\Gamma^0(2)]$ which is $1+\frac{k}{2}$. Also note that
the leading terms of $8\delta_2, \varepsilon_2$ in the lemma have coefficients
$1$ which immediately gives the integrality. The modularity follows from the
transformation formulas of the theta-functions and the fact that $\Gamma^0(2)$
is generated by $STS$ and $T^2STS$.

Now we can {\em prove Theorem 1}. By Lemmas 1 and 2 we can write
$$P_2(\tau)=b_0(8\delta_2)^{2k+1}+b_1(8\delta_2)^{2k-1}\varepsilon_2+\cdots
+b_k (8\delta_2)\varepsilon_2^k$$
where the $b_j$'s are integral linear combinations of the
$\hat{A}(M)\mbox{ch}B_j$'s.

Apply the modular transformation $S:\ \tau\rightarrow -\frac{1}{\tau}$, we have
\begin{eqnarray*}
\delta_2(-\frac{1}{\tau})=\tau^2 \delta_1(\tau)&,&
\varepsilon_2(-\frac{1}{\tau})=\tau^4\varepsilon_1(\tau),\\
P_2(-\frac{1}{\tau})&=&2^{-l}\tau^{4k+2}P_1(\tau).
\end{eqnarray*}
Therefore$$P_1(\tau)=2^l[b_0(8\delta_1)^{2k+1}+b_1(8\delta_1)^{2k-1}\varepsilon_1+\cdots
+b_k (8\delta_1)\varepsilon_1^k].$$
At $q=0$, $8\delta_1=2$ and $\varepsilon_1=2^{-4}$. One gets the result by a
simple manipulation. The cancellation formula for dimension $8k$ case can be
proved in the same way. $\spadesuit$

One can certainly get more divisibility results of characteristic numbers from
the above formulas.

{\em Corollary 1} is an easy consequence of Theorem 1. Since in the case that
$M$ and $V$ are spin, all of the $b_j$'s are integral linear combinations of
the $\mbox{Ind}D\otimes B_j$'s. {\em Corollary 2} follows from Corollary 1,
since in dimension $8k+4$, each $b_j$ is an even integer.

For Corollary 3, we first recall the definition of $\eta$-invariant. Let $(N,
w)=\partial (M, W)$ be as in Section 2. Let $E$ be a real vector bundle on $M$.
Consider a twisted Dirac operator $D\otimes E$ on $M$. With a suitable choice
of metrics on $M, N $ and $E$, in a neighborhood of $N$ one can write
$$D\otimes E=\sigma( \frac{\partial}{\partial u} +D_N\otimes E|_N)$$where $D_N$
is the Dirac operator on $N$, $E|_N$ is the restriction of $E$ to $N$, $\sigma$
is the bundle isomorphism induced by the symbol of $D\otimes E$ and $u$ is the
parameter in the normal direction to $N$. Let $\{ \lambda_j\}$ be the non-zero
eigenvalues of $D_N\otimes E|_N$ and $h_E$ be the complex dimension of its zero
eigenvectors. Then the $\eta$-invariant associated to $D\otimes E$, which we
denote by $\eta(E)$, is given by evaluating at $s=0$ of the function
$$\eta(s)=\sum_j\mbox{sign}\lambda_j\cdot \lambda_j^{-s}.$$ One has the
following index formula from [APS]
$$\mbox{Ind}D\otimes E=\int_M\hat{A}(M)\mbox{ch}E-\frac{\eta(E)+h_E}{2}.$$For
convenience, we will call $\eta(E)$ the $\eta$-invariant associated to
$D\otimes E$. One should note that $\eta(E)$ and $h_E$ are actually geometrical
invariants of $N$.

For the {\em proof of Corollary 3}, we take $V=TM$ in Theorem 1 and apply this
formula to $L(M)$ and each $b_j=\hat{A}(M)\mbox{ch}\beta_j$. For $\beta_j$ the
[APS] formula gives us
$$\mbox{Ind}D\otimes
\beta_j=\int_M\hat{A}(M)\mbox{ch}\beta_j-\frac{\eta(\beta_j)+h_{\beta_j}}{2}.$$For $L(M)$ we have
$$\mbox{sign}(M)=\int_M L(M)-\eta(\triangle)$$ where $\eta(\triangle)$ is the
$\eta$-invariant associated to the signature operator $d_s$. Put these two
formulas into the equality of Theorem 1 with $V=TM$, we get
$$R(N, w)\equiv -\eta(\triangle)+\sum_{j=0}^k
2^{6k-6j+2}(\eta(\beta_j)+h_{\beta_j})(\mbox{mod}16).$$
Here we have used the fact that $\mbox{Ind}D\otimes \beta_j$ is even in
dimension $8k+4$.
\section{Generalizations}
Let $M$ and $V$ be as in Section 2. We introduce two more elements in
$K(M)[[q^{\frac{1}{2}}]]$.
\begin{eqnarray*}
\Phi_1(M,V)&=&\Theta_1(M,TM)\otimes\otimes_{n=1}^\infty\Lambda_{q^n}(V-\mbox{dim}V),\\
\Phi_2(M,V)&=&\Theta_2(M,TM)\otimes\otimes_{n=1}^\infty\Lambda_{-q^{n-\frac{1}{2}}}(V-\mbox{dim}V)
\end{eqnarray*}where $\Theta_j(M,TM)$ is defined as in Section 2 with $V=TM$.

Similarly introduce two cohomology classes on $M$:
\begin{eqnarray*}
Q_1(\tau)&=&L(M)\mbox{ch}\triangle(V)\mbox{ch}\Phi_1(M,V),\\
Q_2(\tau)&=&\hat{A}(M)\mbox{ch}\Phi_2(M,V).
\end{eqnarray*}
Express them in terms of the theta-functions, we have
\begin{eqnarray*}
Q_1(\tau)&=&2^{l+4k+2}(\prod_{j=1}^{4k+2}x_j\frac{\theta'(0,\tau)\theta_1(x_j,\tau)}{\theta(x_j,\tau)\theta_1(0,\tau)})
\prod_{\nu=1}^l\frac{\theta_1(y_\nu,\tau)}{\theta_1(0,\tau)},\\
Q_2(\tau)&=&(\prod_{j=1}^{4k+2}x_j\frac{\theta'(0,\tau)\theta_2(x_j,\tau)}{\theta(x_j,\tau)\theta_2(0,\tau)})\prod_{\nu=1}^l\frac{\theta_2(y_\nu,\tau)}{\theta_2(0,\tau)}.
\end{eqnarray*}
If $p_1(V)=0$, then it is easy to see that the top degree terms of these two
classes are modular forms over $\Gamma_0(2)$ and $\Gamma^0(2)$ respectively.
Similar method to the proof of Theorem 1 can be used to derive an expression of
$L(M)\mbox{ch}\triangle(V)$ in terms of the integral linear combination of the
$\hat{A}(M)\mbox{ch} D_j$'s where the $D_j$'s are the Fourier coefficients of
$\Phi_2(M,V)$. More explicitly we have
$$L(M)\mbox{ch}\triangle(V)=2^{l+3}\sum_{j=0}^k2^{6k-6j}d_j$$where each $d_j$
is the integral linear combination of the $\hat{A}(M)\mbox{ch}D_j$'s. As a
corollary, one gets
\begin{cor}
Let $V$ be a rank $2l$ spin vector bundle on a spin manifold $M$ of dimension
$8k+4$, if $p_1(V)=0$, then $\mbox{Ind}\
d_s\otimes\triangle(V)\equiv0(\mbox{mod}\ 2^{l+4})$.
\end{cor}
We remark that this corollary can also be derived from Corollary 2 by taking
$V\oplus TM$ as the $V$ there.

Using Corollary 3, we can relate the Rochlin invariant to the holonomy of
certain determinant line bundles.
This generalizes the formula in [LMW] to higher dimension.  We first recall
some basic notations.

Given a smooth family of $8k+2$-dimensional compact smooth spin manifolds $\pi
:\ Z\rightarrow N$ and a spin vector bundle $E$ on $Z$, let $D_x$ be the Dirac
operator on the fiber $M_x=\pi^{-1}(x)$. To each $x$ we associate the one
dimensional complex vector space $$(\Lambda^{\mbox{max}} \mbox{ker}D_x\otimes
E|_{M_x})^*\otimes\Lambda^{\mbox{max}} \mbox{coker}D_x\otimes E|_{M_x}$$which
patches together to give a well-defined smooth complex line bundle on $N$,
called determinant line bundle and denoted by $\mbox{det}D\otimes E$. When $E,
N$ and $Z$ are equipped with smooth metrics, then one has the following
equality of differential forms [BF]
$$ c_1(\mbox{det}D\otimes E)_Q=\int_{M_x}\hat{A}(M_x)\mbox{ch}E$$where the left
hand side is the first Chern form with respect to the Quillen metric [BF].

 Let $M$ be a dimension $8k+2$ compact smooth spin manifold, and $f$ be a
diffeomorphism of $M$. Denote by $(M\times S^1)_f$ the mapping torus of $f$.
Recall that it is defined to be
$$M\times [0,1]/(x, 0)\sim(f(x),1).$$ Each spin structure $w$ on $M$ naturally
induces a spin structure which we still denote by $w$, on the mapping torus
[LMW].

Let $\gamma:\ S^1\rightarrow N$ be an immersed circle in $N$. Pulling back from
the family $\pi:\ Z\rightarrow N$, we get a $8k+3$ dimensional manifold which
is isomorphic to a mapping torus $(M\times S^1)_f$. Let $H(E,w)$ denote the
holonomy of the line bundle $\gamma^*\mbox{det}\ D\otimes E$ with spin
structure $w$ on $M$ around the circle. Scaling the (induced) metric on $S^1$
by $\varepsilon^{-2}$ and let $\varepsilon $ go to zero, we have the following
holonomy formula [BF]
$$H(E,w)=\mbox{lim}_{\varepsilon\rightarrow 0}\mbox{exp}(-\pi
i(\eta(E)+h_E)).$$

Given two spin structures $w_1, w_2$ on $M$, From Corollary 3 we have
$$R((M\times S^1)_f, w_1)- R((M\times S^1)_f, w_2)=$$
$$\sum_{j=0}^k 2^{6k-6j+2}\{ (\eta^{w_1}(\beta_j)+h_{\beta_j}^{w_1})-
(\eta^{w_2}(\beta_j)+h_{\beta_j}^{w_2})\}$$where the superscript $w_j$ denotes
the invariant of the corresponding spin structure. Therefore the above holonomy
formula gives us the following
\begin{cor}
$$\mbox{exp}\{ -\frac{\pi i}{4}(R((M\times S^1)_f, w_1)- R((M\times S^1)_f,
w_2))\} =$$
$$\prod_{j=0}^k(H(\beta_j, w_1)H(\beta_j, w_2)^{-1})^{2^{6k-6j}}.$$
\end{cor}
Note that in this equality, the $\beta_j$'s are considered as elements in
$K((M\times S^1)_f)$ by restriction. More precisely they are the restriction to
$(M\times S^1)_{f}$ of the corresponding $\beta_j$'s on $W$ where $\partial
W=(M\times S^1)_f$.

For any oriented compact smooth manifold $M$ with $p_1(M)=0$, the top degree
term of
$$\hat{A}(M)\mbox{ch}\otimes_{n=1}^\infty S_{q^n}(TM-\mbox{dim}M)$$ is a
modular form over $SL_2(Z)$ of weight $k=\frac{1}{2}\mbox{dim}M$. When $M$ is
spin with $p_1(M)=0$, this cohomology class is the index density of the Dirac
operator on the loop space $LM$. See [W], [Liu1] for the other aspects of this
operator. From elementary theory of modular forms, we can easily get the
followiing
\begin{cor}
If $\hat{A}(M)=0$, $p_1(M)=0$ and $\mbox{dim} M< 24$, then the top degree term
of
$$\hat{A}(M)\mbox{ch}\otimes_{n=1}^\infty S_{q^n}TM$$
vanishes.
\end{cor}
This corollary applies to the case that $M$ is spin with positive scaler
curvature or with $S^1$-action. It is interesting to find out the geometric
meaning of this corollary.

Similarly one can consider
\begin{eqnarray*}
\hat{A}(M)&\mbox{ch}&(\triangle^+(M)-\triangle^-(M))
\otimes_{n=1}^\infty S_{q^n}(TM-\mbox{dim}M)\\
&\mbox{ch}&\otimes(\triangle^+(V)-\triangle^-(V))\otimes_{m=1}^\infty\Lambda_{q^m}(V-\mbox{dim}V)
\end{eqnarray*}
which is the index density of the Euler characteristic operator on $LM$. Here
$$\mbox{ch}(\triangle^+(V)-\triangle^-(V))=\prod^l_{j=1}(e^{\pi iy_j}-e^{-\pi
iy_j})$$ and $\mbox{ch}(\triangle^+(M)-\triangle^-(M))$ denotes the
corresponding class for $TM$.

If $p_1(V)=p_1(M)$, its top degree term is a modular form of weight $k-l$ over
$SL_2(Z)$. Here $2l=\mbox{rank}V$ and $2k=\mbox{dim}M$. When $l>k$ or $k=l$ and
the Euler characteristic of $V$ is zero, the top degree term of this class
vanishes.
\section{Discussions}
Modular invariance is one of the most fundemental principle in modern
mathematical physics. In [Liu1], we proved that many elliptic operators derived
from loop space have modular invariance which in turn implies their rigidity.
We note that formal application of the Lefschetz fixed point formula on loop
space together with modular invariance gives us a lot of information about the
index theory on loop space. Many topological results, such as rigidity and
divisibility, are much easier to understand when we look at them while standing
on loop space, therefore they are infinite dimensional phenomena. On the other
hand many seemingly unrelated topological results discussed above are also
intrinsically connected together by the modular invariance of elliptic
operators derived from loop space.

Another interesting point is not directly related to modular invariance. It is
only a kind of wild speculation. As observed by Witten, the inverse of the
$\hat{A}$-genus is actually the equivariant Euler class of the normal bundle of
$M$ in its loop space $LM$ with respect to the natural $S^1$-action. Therefore
up to certain normalization, the Atiyah-Singer index formula can be formally
derived from the Duistermaat-Heckman localization formula on $LM$. See [A]. We
note that, up to certain normalization, the inverse of the loop space
$\hat{A}$-genus
$$q^{-\frac{k}{12}}\hat{A}(M)\mbox{ch}\otimes\otimes_{n=1}^\infty
S_{q^n}TM$$where $2k=\mbox{dim}M$, is formally equal to the equivariant Euler
class of the normal bundle of $M$ in the double loop space $LLM$ with respect
to the natural $T=S^1\times S^1$-action. Up to certain normalization constant,
this is a corollary of the following 'product' formula
$$\prod_{m,n}(x+m\tau+n)=q^{\frac{1}{12}}(e^{\pi ix}-e^{-\pi
ix})\prod_{n=1}^\infty(1-e^{2\pi ix}q^n)(1-e^{-2\pi ix}q^n).$$ This means that
the index formula on loop space, especially elliptic genera, can also be
derived from the formal application of the Duistermaat-Heckman formula to the
double loop space $LLM$!

More precisely, one knows that the $K$-theory Euler class of $TM$ is
$\triangle^+(M)-\triangle^-(M)$, while under the Chern character map, we have
$$\mbox{ch}\ ( \triangle^+(M)-\triangle^-(M))=\mbox{the equivariant Euler class
of} \ TLM|_M$$
where $TLM|_M$ denotes the restriction of the tangent bundle of $LM$ to $M$.
Similarly the equivariant $K$-theory Euler class of $TLM|_M$ is
$$q^{\frac{k}{12}}(\triangle^+(M)-\triangle^-(M))\otimes_{n=1}^\infty
\Lambda_{-q^n}TM$$and its Chern character is
$$q^{\frac{k}{12}}(e^{\pi ix}-e^{-\pi ix})\prod_{n=1}^\infty(1-e^{2\pi
ix}q^n)(1-e^{-2\pi ix}q^n)$$which is also the equivariant cohomology Euler
class of $TLLM|_M$ with respect to the natural $T$-action.

Let us use $\Leftrightarrow$ to mean 'correspondence', then modulo certain
normalization, we can put the above discussions in the follwoing way.
\begin{center}
The equivariant cohomology Euler class of $TLM|_M\Leftrightarrow$

the $K$-theory Euler class of $TM$;

The equivariant cohomology Euler class of $TLLM|_M\Leftrightarrow$

 the equivariant $K$-theory Euler class of $TLM|_M$
\end{center}

Another interesting point is that, under the natural transformation of
cohomology theory
$$\mbox{Ell}^*(M)\rightarrow K(M)[[q]],$$
the Euler class of elliptic cohomology is transformed to the equivariant
$K$-theory Euler class of $TLM|_M$.
We remark that the elliptic cohomology here may be slightly different from the
one in [L]. Actually the elliptic cohomology in [L] is associated to the
signature operator on loop space, while the one here should be associated to
the Dirac operator on loop space, which conjecturally should exist.

These discussions make it reasonable to speculate that the equivariant
cohomology of $LM$ should correspond to the $K$-theory of $M$, while the
equivariant cohomology of the double loop space $LLM$ should correspond to the
equivariant $K$-theory of $LM$ which in turn corresponds to the elliptic
cohomology of $M$. We can put these into the following diagram
\begin{eqnarray*}
H_{S^1}(LM)&\Leftrightarrow& K(M)\\
H^*_{T}(LLM)&\Leftrightarrow& K_{S^1}(LM),\\
K_{S^1}(LM)&\Leftrightarrow&\mbox{Ell}^*(M)\Leftrightarrow H^*_T(LLM).
\end{eqnarray*}
This means that 'looping' $M$ once 'lifts' the equivariant cohomology theory
one order higher.

Department of Mathematics

MIT

Cambridge, MA 02139
\end{document}